\documentclass[twocolumn,twoside,slac_two]{revtex4}
\usepackage{graphicx}
\usepackage{fancyhdr}
\newcommand{\g}{$\gamma$}

\newcommand{\Hii}{H\rmnum{2}}

\newcommand{\sdeg}{$^{\circ}$}                         
\newcommand{\ddeg}{\hbox{$.\!\!^\circ$}}               
\newcommand{\TS}{{$TS$ }}

\makeatletter
\newcommand{\rmnum}[1]{\romannumeral #1}
\newcommand{\Rmnum}[1]{\expandafter\@slowromancap\romannumeral #1@}
\makeatother

\begin{document}

\title{Fermi LAT gamma-ray observations of the supernova remnant HB~21}

\author{G.Pivato}
\affiliation{University and INFN Padova, via Marzolo 8, 35131 Padova, Italy}

\author{J.W.Hewitt}
\affiliation{NASA Goddard Space Flight Center, Greenbelt, MD 20771, USA}

\author{L.Tibaldo}
\affiliation{KIPAC - SLAC, MS 29,2575 Sand Hill Rd., Menlo Park, CA 94025, USA}
\author{on behalf of Fermi LAT collaboration}

\begin{abstract}
We present the analysis of Fermi Large Area Telescope (LAT) \g-ray observations
of HB~21, a mixed-morphology supernova remnant. Such supernova remnants are
characterized by an interior thermal X-ray plasma, surrounded by a wider
nonthermal shell emitting at radio frequencies. HB~21 has a large angular size,
making it a good candidate for detailed morphological and spectral studies with
the LAT. The radio extension is $2^\circ\times$1\ddeg5, compared to the LAT
$68\%$ containment angle of $\sim1^\circ$ at 1~GeV. To understand the origin of
\g-ray emission, we compare LAT observations with other wavelengths that trace
non-thermal radio synchrotron, nearby molecular clouds, shocked molecular
clumps, and the central X-ray plasma. Finally, we model possible hadronic and
leptonic emission mechanisms. We conclude that \g-rays from HB~21 are likely the
result of electron bremsstrahlung or proton-proton collisions with dense
material due to interaction with the nearby clouds.
\end{abstract}

\maketitle

\thispagestyle{fancy}

\section{INTRODUCTION}

HB~21 (also known as G89.0+4.7, \cite{green}) is a mixed morphology (MM) supernova remnant (SNR) interacting with molecular clouds. X-rays observations with \emph{Einstein} provide evidence that the supernova (SN) explosion took place in a low-density cavity, therefore suggesting a massive stellar progenitor (\cite{kno96}). HB~21 is located in a dense environment with a radio shape suggestive of interactions with the nearby clouds (\cite{Byu06}). Looking at the CO distribution it is possible to distinguish two different structures based on the velocity of the molecular clouds. \cite{koo91} report the presence of broad CO lines
emitted from shocked clumps of molecular gas with high column densities.
The distance to HB~21 was first estimated to be $0.8$~kpc from
ROSAT X-rays observations (\cite{lea96}). Considering the preshock velocities,
the X-ray absorbing column densities, the relation between photometric distance
of \Hii~region and velocity of molecular clouds and the highly polarized
emission, the distance was later determined to be $\sim$1.7 kpc (\cite{Byu06}). 

High-energy \g-ray sources coincident with HB~21 were reported in the 2-years
Catalog of the Large Area Telescope (LAT) on board the \emph{Fermi Gamma-ray
Space Telescope} (\cite{2fgl}). Three sources are associated with the remnant
(2FGL~J2043.3+5105, J2046.0+4954, and J2041.5+5003) and an additional source
lies on the edge (2FGL~J2051.8+5054). Due to the large apparent size
($\sim$2\sdeg) this object is well suited for a detailed morphological study
with \emph{Fermi} LAT. Here we present the analysis of $\sim4$~years of LAT
observations of HB~21 and the discussion of the \g-ray emission mechanism in
light of archival multiwavelength observations. In \S\ref{analysis} we describe
our morphological and spectral characterization of the \g-ray emission from
HB~21, the sources of systematic errors and the strategy adopted to extimate
them. In \S\ref{discussion} the interpretation of these results is finally
discussed. This paper is intended to be a summary of the poster we presented at
Fermi Symposium. For a detailed description of the analysis and a more
comprehensive discussion of the results we defer the reader to the paper in
preparation \cite{hb21pap}.

\section{\label{analysis}FERMI LAT DATA ANALYSIS}

For this analysis, we use data accumulated from the beginning of scientific
operations, on 2008 August 4, to 2012 June 14, selecting the low-background
event class P7SOURCE$\_$V6. For the morphological characterization we
use only events with energy $>$1~GeV to profit from the narrower point-spread
function (PSF) in order to separate the \g-ray emission associated with HB~21
from neighboring sources and interstellar emission. We then use events down to
$100$~MeV to determine the spectral energy distribution of the remnant.
Below this threshold the PSF becomes much broader than the SNR and the uncertainties related to the instrument response are larger. In both the morphological and spectral characterization we consider photons with measured energies up to $300$~GeV, but only find a significant detection of the source up to energies of several GeV due to the limited number of events at high energies.

We perform the
analysis in a 10\sdeg$\times$10\sdeg\ region of interest (RoI)
centered at the radio position of HB~21 ($l$=89\ddeg0, $b$=+4\ddeg7).
The background is composed of diffuse emission and individual nearby \g-ray
sources. Diffuse emission is taken into account using the standard models
provided by the \emph{Fermi} LAT
collaboration.
We include in the background model all the point sources present in the 2FGL
catalog \cite{2fgl} with distances less than 15\sdeg\ from the RoI center and
which are not associated with the SNR. We will discuss in \S\ref{morphology} the
case of the source 2FGL~J2051.8+5054, which is located at the edge of the SNR.
The spectral model used for background sources is that reported in the 2FGL
catalog. Fluxes and spectral indices are left as free parameters in the fit if the source is within the RoI. Otherwise they are fixed to the catalog values.

Two major sources of systematic errors on the results are the uncertainties in
the LAT effective area and the modeling of interstellar emission. The
uncertainties in the effective area for the IRFs we use are evaluated as $10\%$
at $100$~MeV, $5\%$ at $516$~MeV, and $10\%$ above $10$~GeV, linearly varying
with the logarithm of energy between those values (\cite{bal12}). We estimate
the error induced in the characterization of the \g-ray emission spectrum from
HB~21 by repeating the analysis with two sets of modified IRFs where the
effective area was upscaled or downscaled by its
uncertainty.
To gauge the systematic uncertainties due to the interstellar emission model we
compare the results obtained using the standard model with the results based on
eight alternative interstellar emission models (\citealt{diffsystpap}).

\subsection{\label{morphology}Morphological study}

\begin{figure*}[htbp]
 \centering
\includegraphics[width=0.40\textwidth]{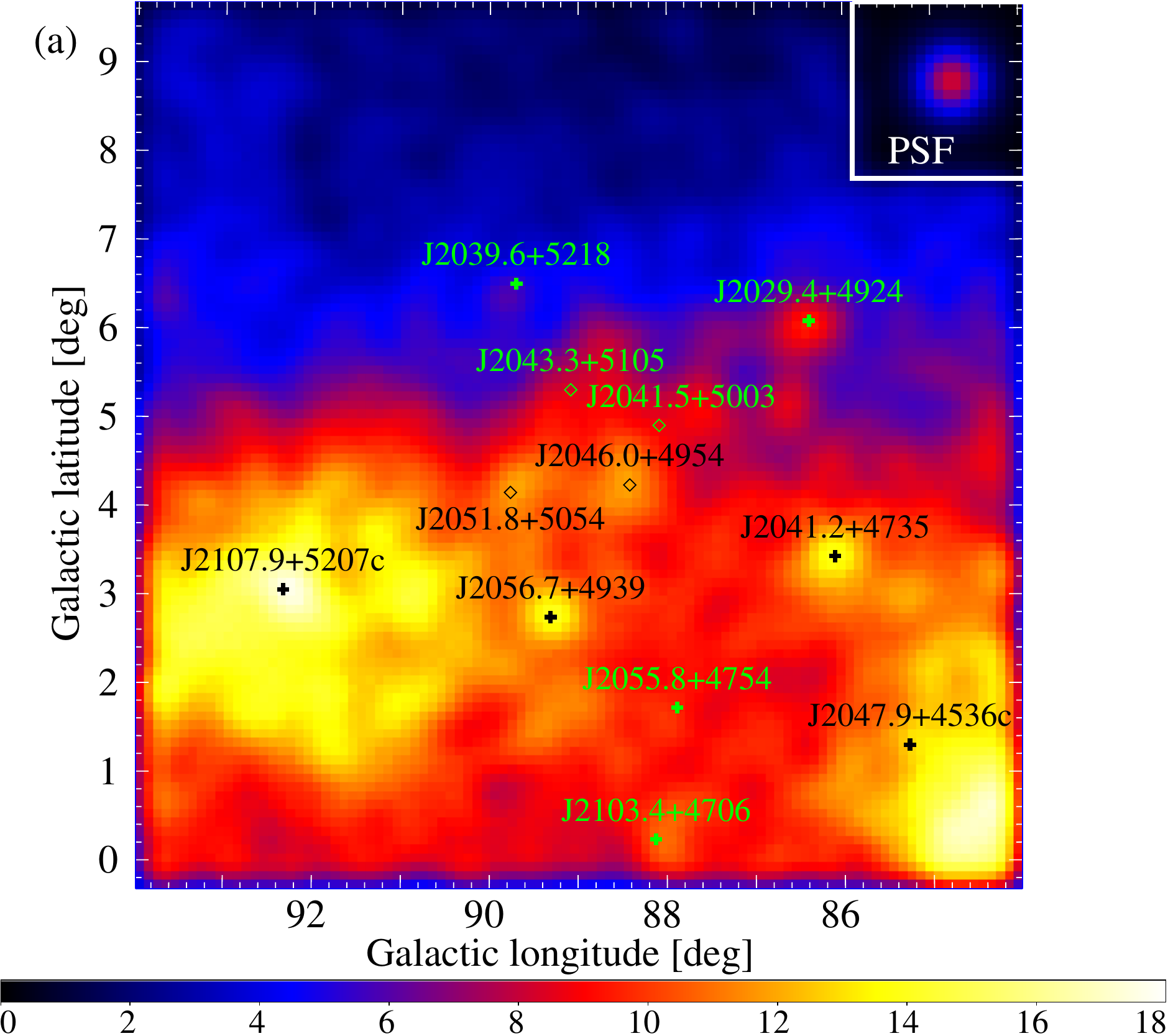}
\put(-230,14){\hbox{\Large{\texttt{\textbf{Preliminary}}}}}
\includegraphics[width=0.40\textwidth]{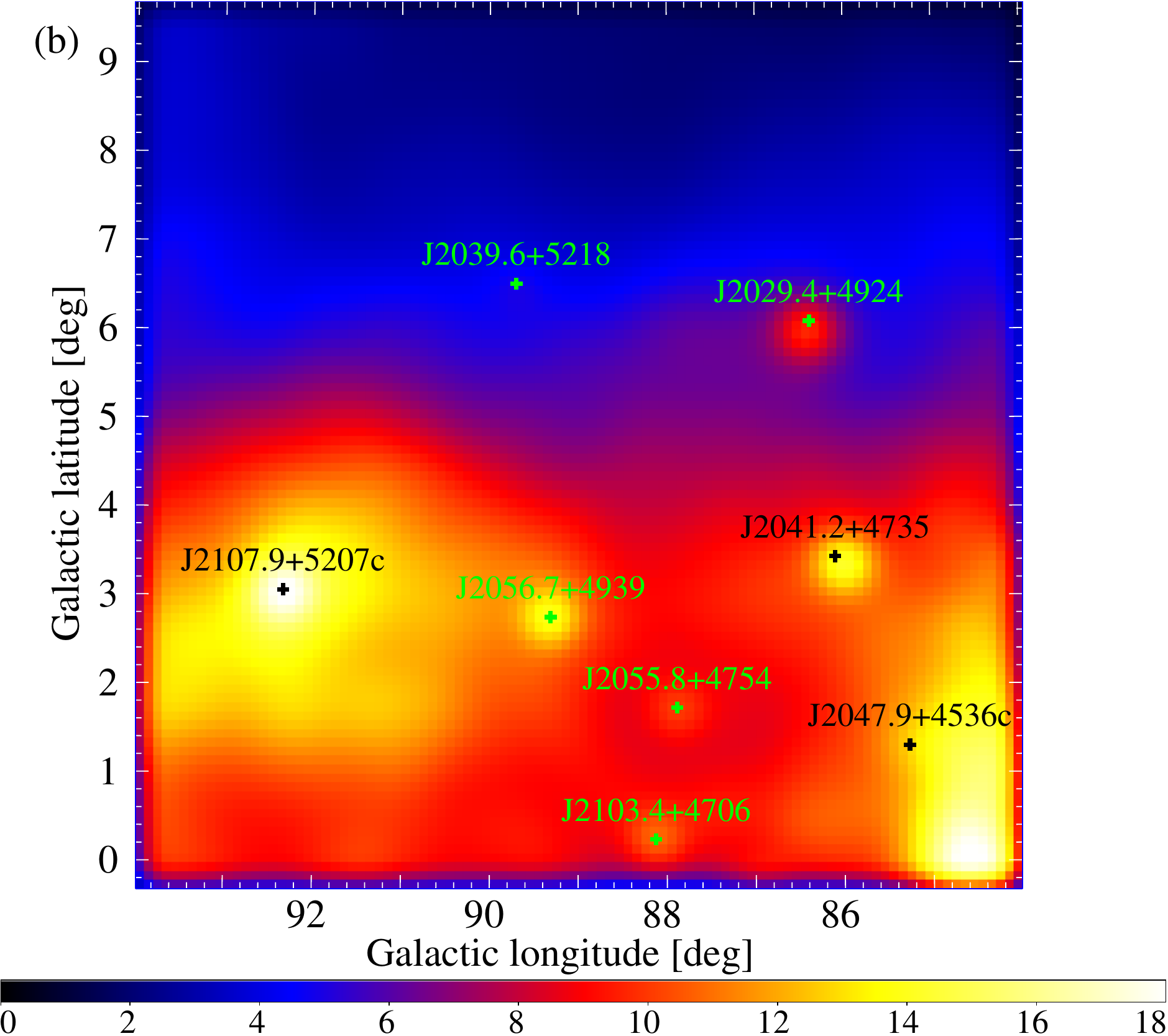}
\put(-50,14){\hbox{\Large{\texttt{\textbf{Preliminary}}}}} \\
\includegraphics[width=0.40\textwidth]{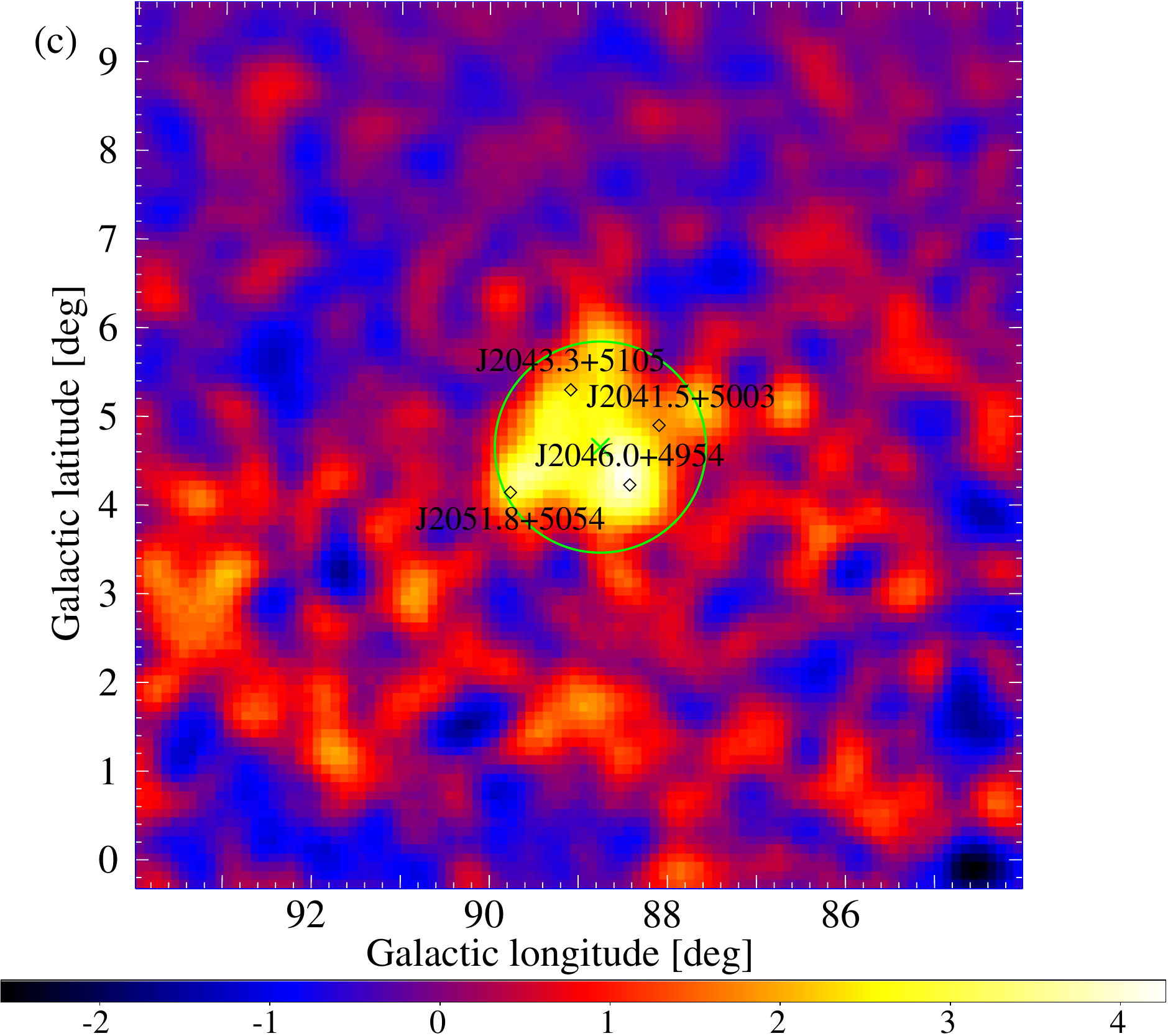}
\put(-50,14){\hbox{\Large{\texttt{\textbf{Preliminary}}}}}
 \caption{\label{maps}a) LAT counts map at energies $> 1$~GeV. We overlaid the positions of 2FGL sources (crosses for background sources and diamonds for the three sources associated with the remnant and 2FGL~J2051.8+5054). The inset in the top right corner shows the effective PSF over the energy range considered for a power-law spectral distribution with index 3.1. 
b) Background model map (calculated using the fit parameters for the case when HB~21 is modeled using the best-fit disk).  
c) Remaining emission associated with HB~21; overlaid are the positions of  the four point sources above and the best-fit disk. The pixel size is 0\ddeg1 and all maps are smoothed for display with a Gaussian kernel of  $\sigma$=0\ddeg4.}
 \end{figure*}

In Figure~\ref{maps}a we show a count map of the RoI for energies $>1$~GeV, to
visually illustrate the morphology of the \g-ray emission in the region. 

\begin{figure*}[htbp]
\centering
\includegraphics[width=0.40\textwidth]{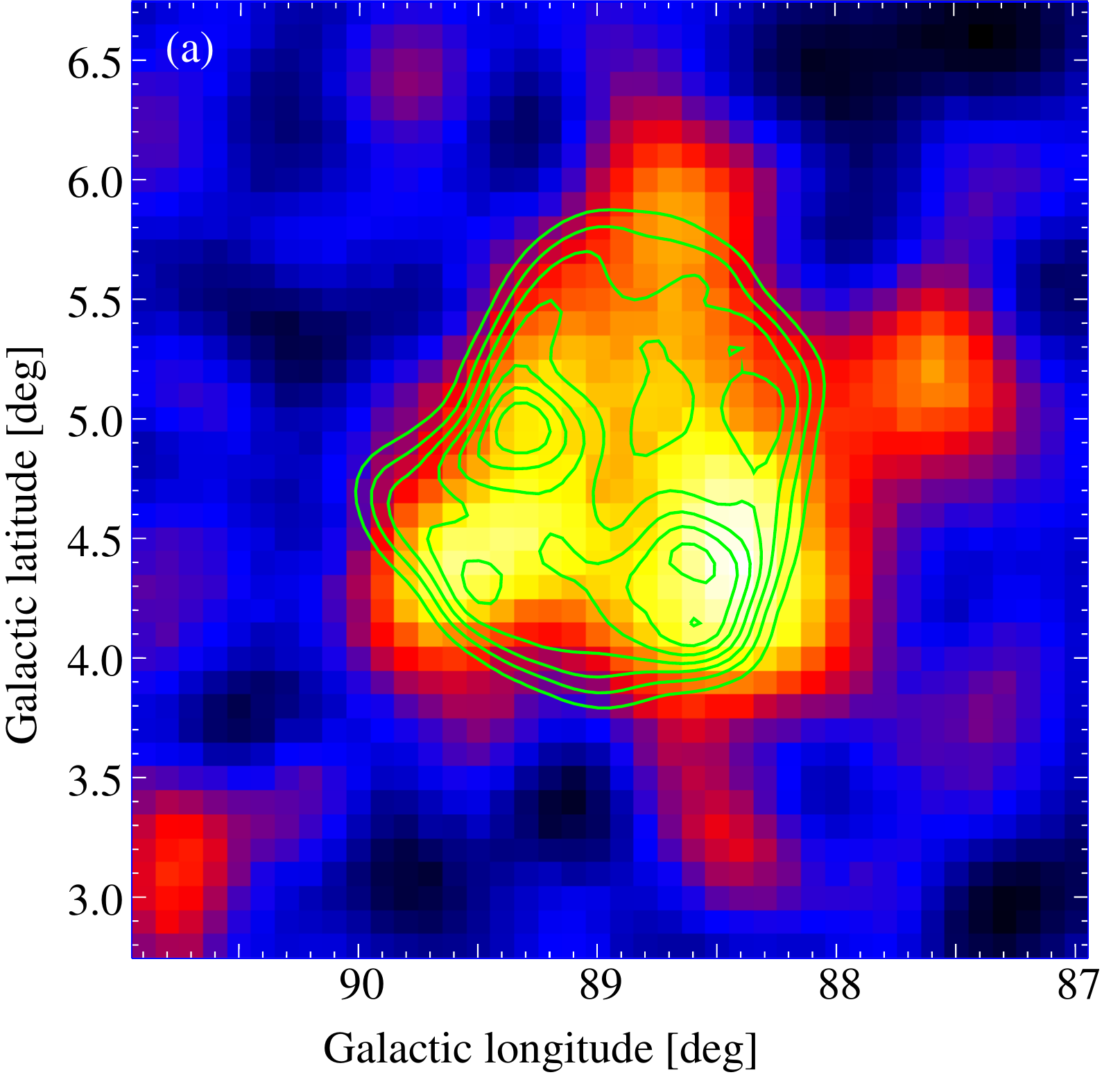}
\put(-230,0){\hbox{\Large{\texttt{\textbf{Preliminary}}}}}
\includegraphics[width=0.40\textwidth]{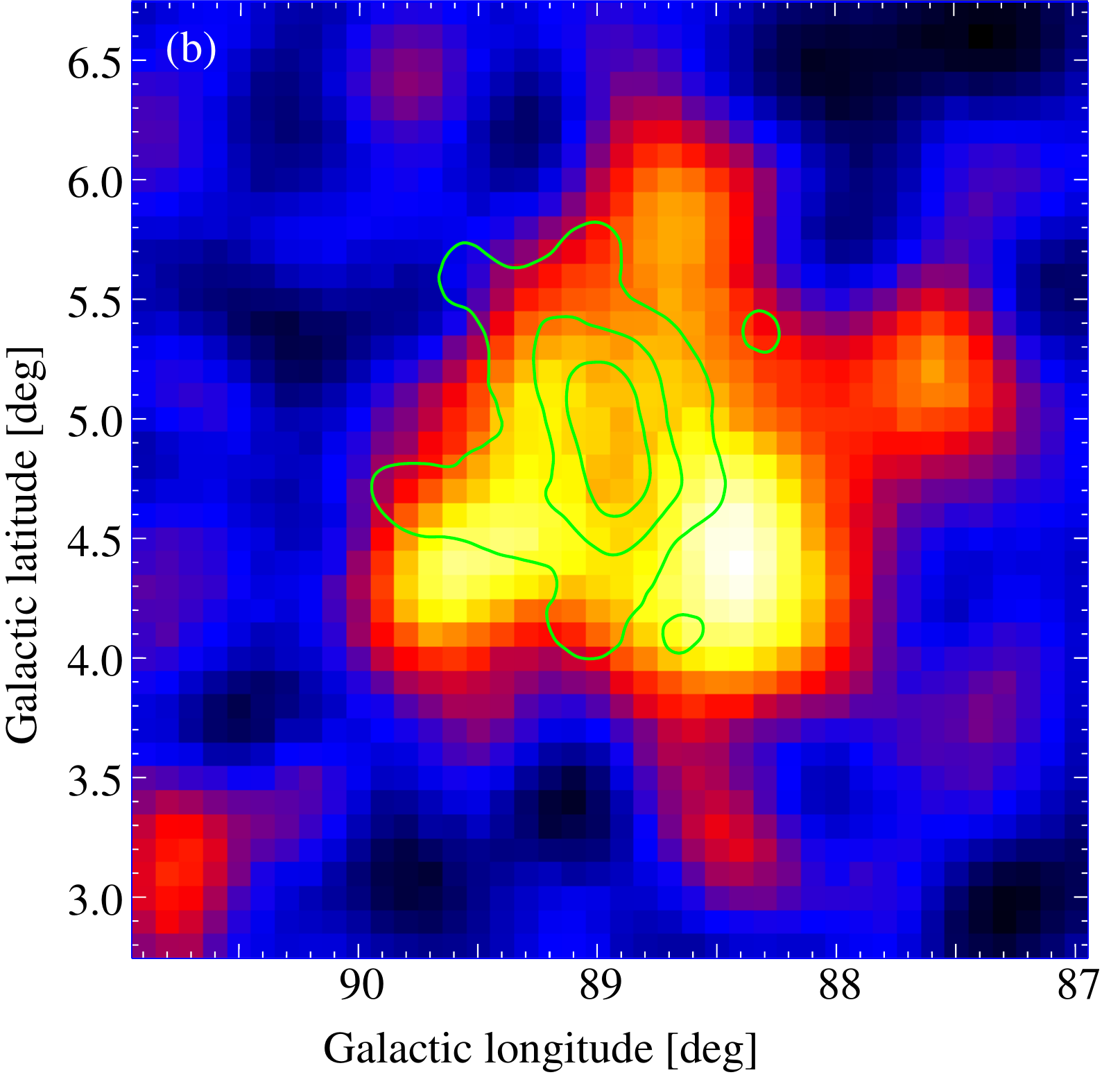} 
\put(-50,0){\hbox{\Large{\texttt{\textbf{Preliminary}}}}}\\
\includegraphics[width=0.40\textwidth]{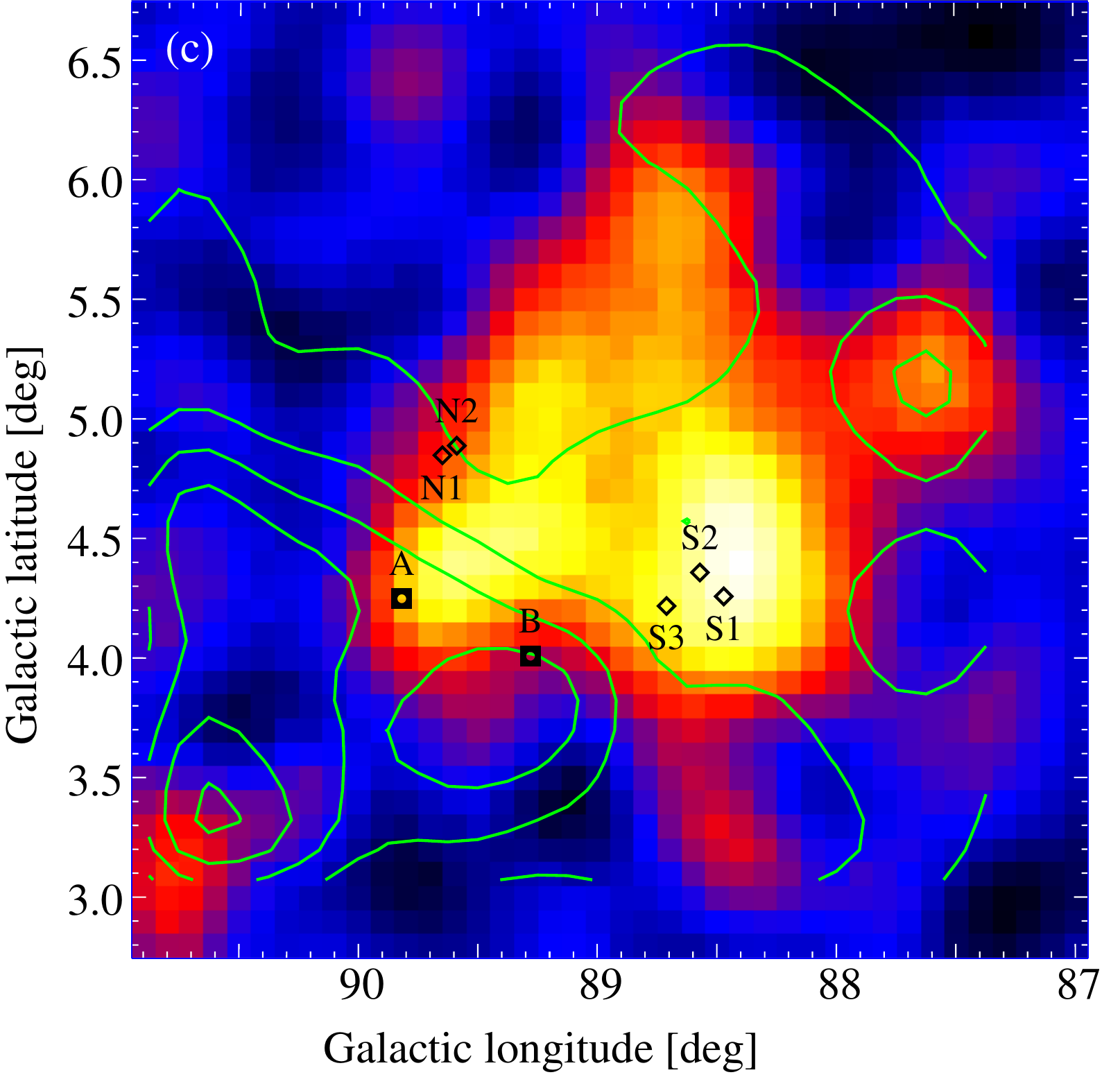}
\put(-50,0){\hbox{\Large{\texttt{\textbf{Preliminary}}}}}\\
\includegraphics[width=0.39\textwidth]{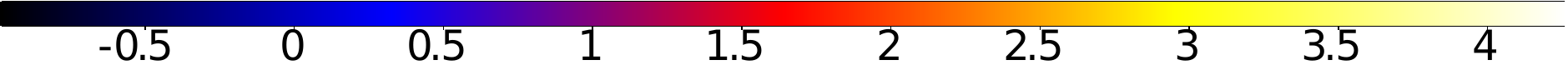}
\caption{\label{over}Emission associated with HB~21 (Fig.~\ref{maps}c) overlaid
with contours from emission at other wavelengths:
a) Radio emission intensity at $6$~cm from Sino-German Telesope \citep{gao11}. Data were smoothed with a Gaussian kernel of
$\sigma$=0\ddeg2. The seven contour levels are linearly spaced from
0~Jy~beam$^{-1}$ to 2.0~Jy~beam$^{-1}$.
b) Background-subtracted X-ray emission smoothed with a Gaussian kernel of
$\sigma$=0\ddeg25. The three contour levels are linearly spaced form
0.36$\times10^{-3}$ to 2.13$\times10^{-3}$
counts~s$^{-1}$~arcmin$^{-2}$.
c) Intensities of the 2.6~mm CO line in the Local-Arm region. The six contour
levels are linearly spaced from 1.5~K~km~s$^{-1}$ to 28~K~km~s$^{-1}$. We also
show the positions of the shocked CO clumps and clouds A and B given in
\cite{koo91}.}
\end{figure*}

We compare the morphology in \g\ rays with that in the radio 
(see figure \ref{over}a) and an X-ray band (see figure \ref{over}b).
As shown in figure \ref{over}b the \g-ray emission is broader than the X-ray (ROSAT) emitting plasma. Figure \ref{over}a shows that \g~emission well compares with radio shell even if it extends beond the radio shell in region where CO clouds are present. Furthermore, note that the brightest part of \g-ray emission coincides with known shocked molecular clouds (labelled S1,S2, and S3 in figure \ref{over}c).

To
characterize the morphology in the \g-ray band we approximate the SNR with a
disk, for which we determine from \g-ray data the best-fit center position and
radius.
We considered both the cases where 2FGL~J2051.8+5054 is included as a separate point source in the model or removed.
The significance of the separate source hypothesis is below the threshold usually required to claim a detection for LAT sources
so we do not consider 2FGL~J2051.8+5054 as a separate source and, therefore, we remove it from the model. The best fit parameter in this case are: $l$=88\ddeg75$\pm$0\ddeg04, $b$=4\ddeg67$\pm$0\ddeg05, $r$=1\ddeg19$\pm$0\ddeg05. Errors reflect statistical uncertainties only. 
We note that the results concerning HB~21 are robust regardless of whether 2FGL~J2051.8+5054 is included as a separate source in the model or not. 

When the alternative interstellar emission models are used, the \g-ray disk is
systematically shifted toward the north-western part with respect to the radio
shell (with shifts in longitude between 0\ddeg19 and 0\ddeg24, and in latitude
between 0\ddeg06 and 0\ddeg09), and the disk radius is systematically
smaller by 0\ddeg18--0\ddeg24. 
We note that, as for the standard model, for all the alternative models the disk
extends beyond the rim of the remnant in coincidence with the western molecular
cloud.

\subsection{\label{spectrum}Spectral study}

Modeling HB~21 with the best-fit disk we determine the spectrum over the
full energy range. 
We study the spectral shape of the \g~emission in the energy range between 100~MeV and 300~GeV. We compare a curved spectrum
\begin{displaymath}
\frac{dN}{dE}=N_0\left(\frac{E}{1000~MeV}\right)^{-(\alpha+\beta\ln(E/1000~MeV))}
\end{displaymath}
 with a power law. We obtain that the curved spetrum is preferred at $9\sigma$, so now on our spectral shape in all energy range will be the log-parabola.
Using the log-parabolic function, the total
energy flux from HB~21 at energies above 100 MeV is
$9.4\pm0.8\;(\mathrm{stat})\pm1.6\;(\mathrm{syst})\times10^{-11}$ erg cm$^{-2}$
s$^{-1}$ and the photon flux 
$1.48\pm0.2\;(\mathrm{stat})\pm0.4\;(\mathrm{syst})\times10^{-11}$ ph cm$^{-2}$
s$^{-1}$.

\begin{figure}
 \centering
 \includegraphics[width=0.5\textwidth]{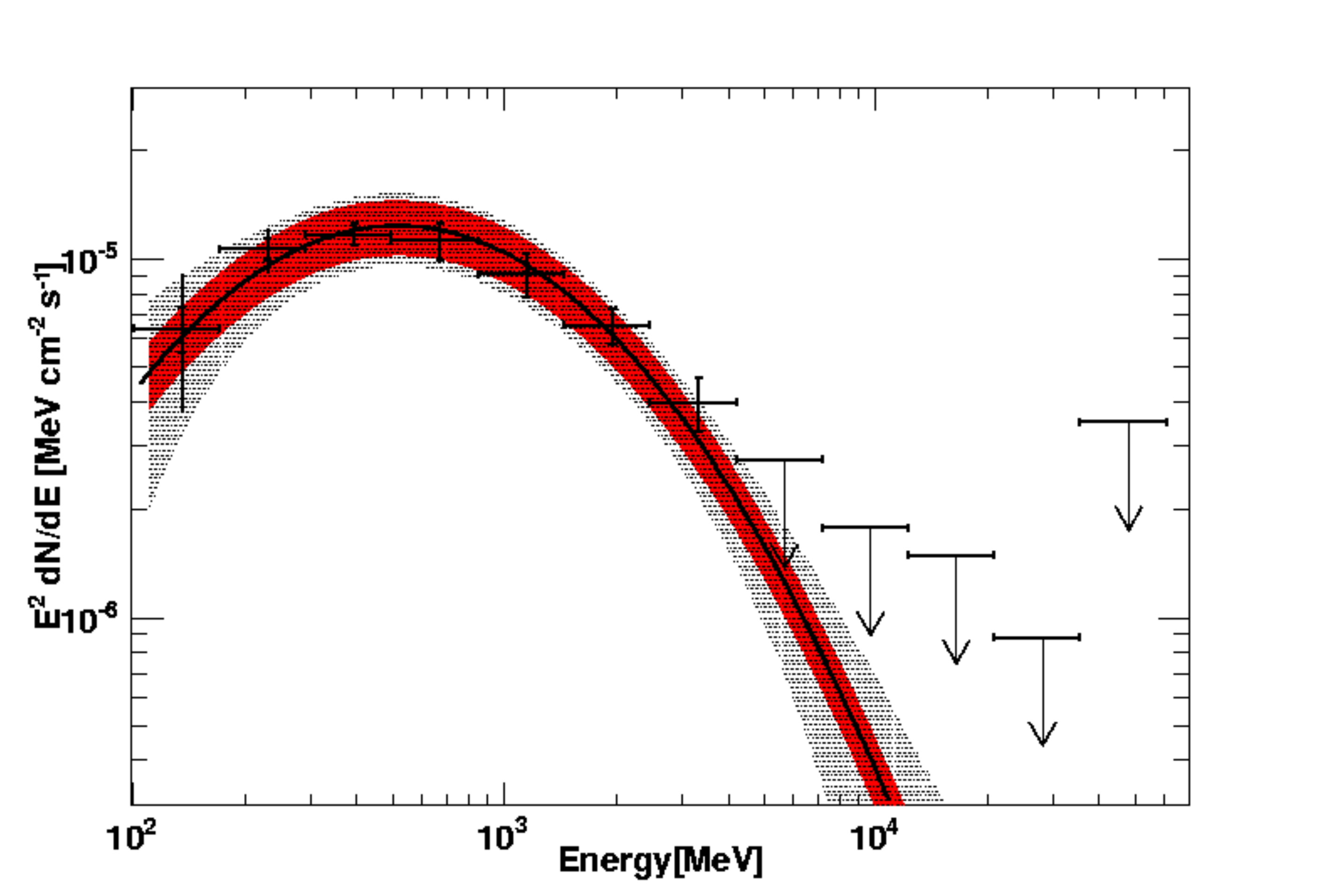}
\put(-130,130){\hbox{\Large{\texttt{\textbf{Preliminary}}}}}
 \caption{\label{sed}Spectral energy distribution (SED) of HB~21. The line shows
the best-fit log-parabola model, the light filled area shows the statistical
error band  and the gray area shows the systematic uncertainties. For the points
error bars correspond to statistical errors only, while lines show the larger
systematic errors. 95\% confidence-level upper limits are given for energy bins
where the \TS of the source is $<9$.}
 \end{figure}

We also computed the spectral energy distribution (SED) in a model-independent
way by splitting the full energy range in 12 logarithm-spaced bins.
We show the resulting SED in Figure~\ref{sed}. Systematic errors related to the
modeling of interstellar emission are obtained repeating the analysis using the
alternative models and extracting the root mean
square of the variations with respect to the values from the standard model.
Those values are then summed in quadrature with the error due to the LAT
effective area uncertainties for display.
At energies below a few hundred MeV the systematic errors are driven by the
uncertainties related to the modeling of interstellar emission because the broad
PSF reduces the source to background ratio. The
effective area uncertainties have a comparable or larger impact at higher
energies.

\section{\label{discussion}SPECTRAL ENERGY DISTRIBUTION MODELING}

To constrain the particle distribution we
simultaneously fit radio (\citet{lea06}) and \g-ray emission from nonthermal
electrons and protons.  We adopt the simplifying assumption that all emission
originates from a region characterized by a constant matter density and magnetic
field strength. This single emitting zone is assumed to be equal to the size of
the remnant derived from the best-fit gamma-ray disk. The particle spectra are
assumed to follow a power-law with an exponential cutoff of the form
$dN/dE\simeq \eta_{e,p}E^{-\Gamma}\times\exp(–E/E_{max})$, with the same
spectral index and energy cutoff for both electrons and protons.

\begin{figure}[htbp]
\centering
\includegraphics[width=0.50\textwidth]{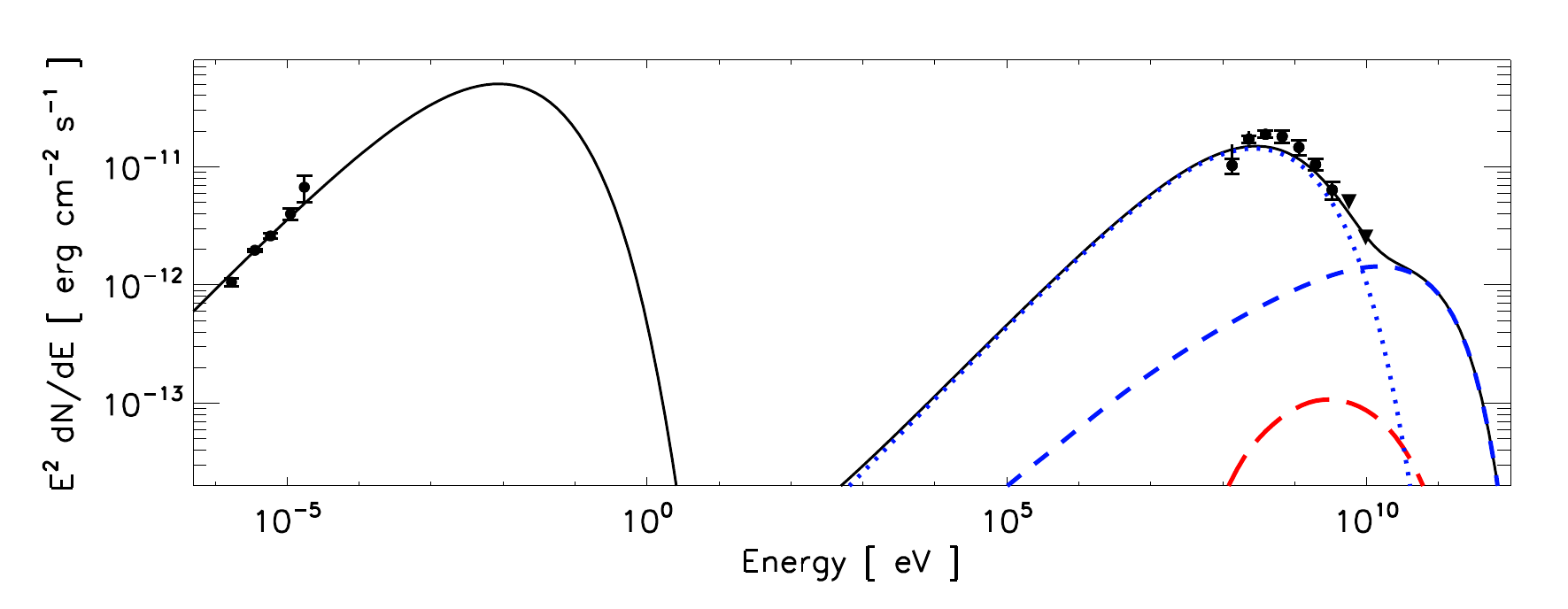} 
\put(-140,70){\hbox{\large{\texttt{\textbf{Preliminary}}}}}\\
\includegraphics[width=0.50\textwidth]{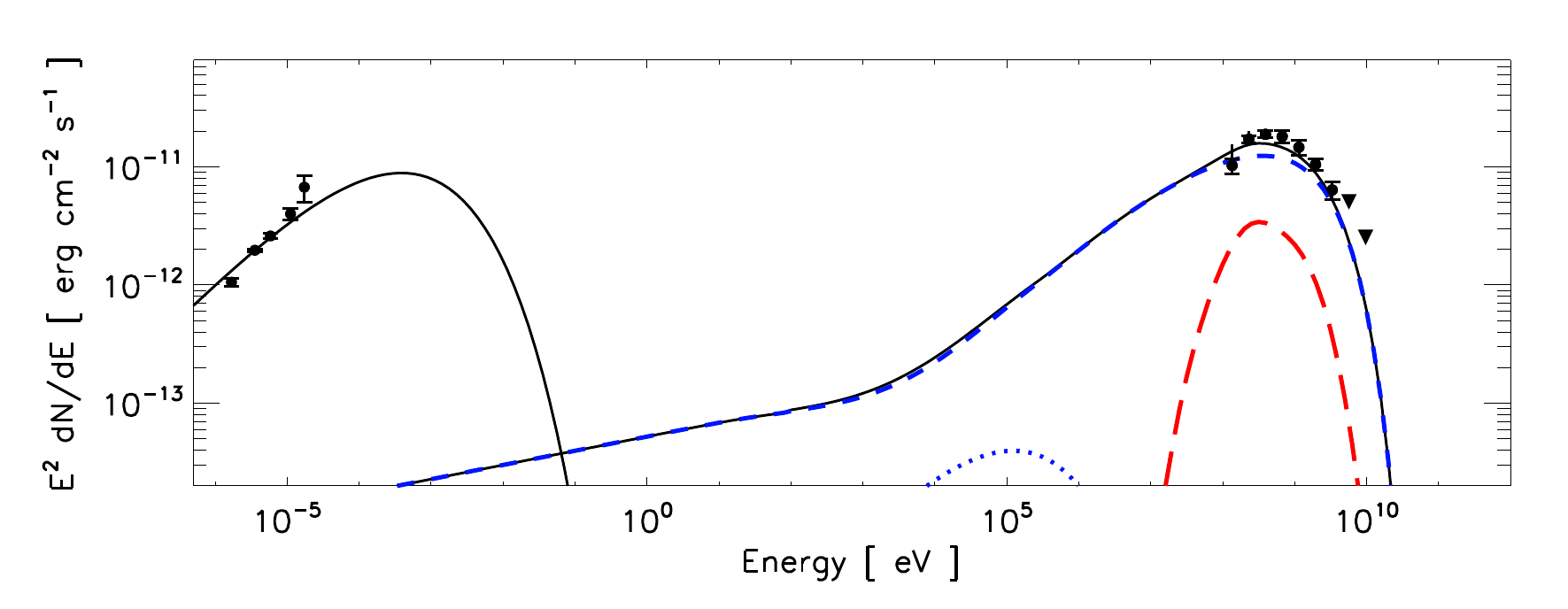} 
\put(-140,70){\hbox{\large{\texttt{\textbf{Preliminary}}}}}\\
\includegraphics[width=0.50\textwidth]{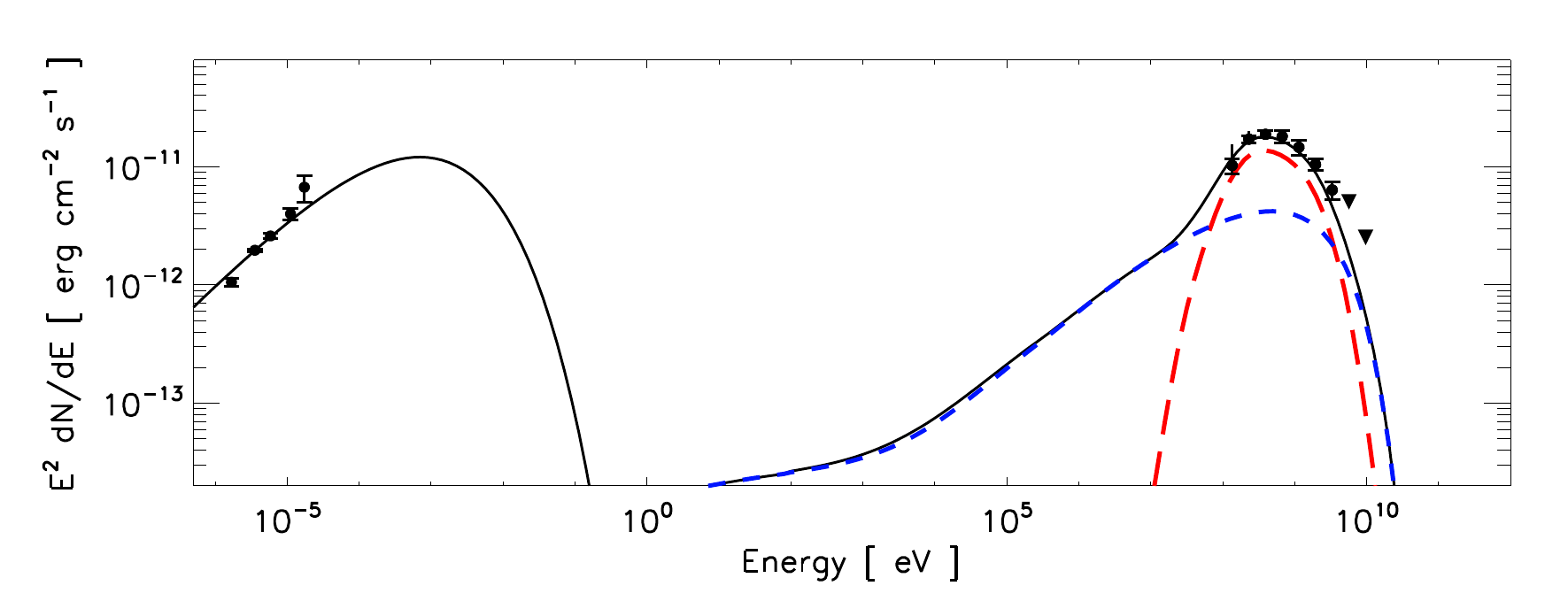} 
\put(-140,70){\hbox{\large{\texttt{\textbf{Preliminary}}}}}\\
\caption{SED models for which IC (top), bremsstrahlung (middle) and $\pi^0$-decay (bottom) are the dominant emission mechanism. In each model the radio data are fit with a synchrotron component, shown as a black curve. The individual contributions of $\pi^0$-decay (long dashed), bremsstrahlung (short dashed), and IC emission from CMB (dotted) are shown. The sum of the \g--ray emission is shown by the solid curve. The leptonic and hadronic components are colored blue and red, respectively.}
\label{sedmod}
\end{figure}

The radio index, $\alpha$, is related to the \g-ray photon index, $\Gamma$, by
$\Gamma=2\alpha+1$ for bremsstrahlung or pion-decay models, giving
$\Gamma=1.76$ (We note that for this particle index, both a cutoff and a
spectral break by an index of one give an equally valid fit). Modeled spectral
energy distributions (SEDs) are presented in \ref{sedmod}. Model parameters are given in Table \ref{tab_sedmod}.
\begin{table}[t]
\begin{center}
\caption{One-Zone Model Parameters}
\begin{tabular}{|c|c|c|c|c|c|c|c|}
\hline 
\textbf{Model} & \textbf{$\Gamma$} & \textbf{$p_{max}$} & \textbf{$n_H$} & \textbf{$B_{tot}$} & \textbf{$\eta_e/\eta_p$} & \textbf{$W_p$} & \textbf{$W_e$} \\
  &   & [GeV/c] & [cm$^{-3}$] & [$\mu$G] &   & [erg] & [erg] \\
\hline 
IC & 1.76 & 200 & 0.1 & 3 & 1 & $1\times10^{49}$ & $4\times10^{49} $\\
\hline 
Brems. & 1.76 & 4 & 10 & 100 & 0.1 & $5\times10^{48}$ & $4\times10^{49}$ \\
\hline
$\pi^0$-decay & 1.76 & 6 & 10 & 100 & 0.01 & $2\times10^{49}$ & $1\times10^{48}$ \\
\hline
\end{tabular}
\label{tab_sedmod}
\end{center}
\end{table}
We also give the total energy of accelerated particles integrated above 1 GeV
for protons, and above 511 keV for electrons. While the chosen parameters are
not unique in their ability to fit the broadband spectrum, they are
representative.

The energetics implied by nonthermal models are in line with
those determined with other old SNRs in a dense environment. Regardless of the
dominant emission mechanism, the energetic requirement is for
$\sim$few$\times10^{49}$ ergs. The hadronic model indicates $\sim2\times10^{49}$
erg in accelerated cosmic ray protons and nuclei. Leptonic models require
$\sim4\times10^{49}$ ergs in electrons. SNR HB 21 is known to be interacting
with molecular clouds, and a moderately enhanced average density would favor
hadronic of bremsstrahlung models for \g-ray emission. The IC-dominated model
requires a low density ($N_H\leq0.1$ cm$^{-3}$) to prevent bremsstrahlung
emission from dominating. 
Such a density is unreasonably low for HB 21.  Interestingly, HB 21 has an
uncharacteristically low luminosity compared to other \g-ray SNRs known to be
interacting with molecular clouds. The total luminosity of the HB 21 above 100
MeV at a distance of 1.7 kpc is $4\times10^{34}$ erg s$^{-1}$, while most
interacting SNRs, such as IC 443 have luminosities of $10^{35}$ erg s$^{-1}$ or
greater. It is possible that the lower luminosity of HB 21 is due to a lower
total mass of the molecular clouds with which it is known to be interacting.

\section{CONCLUSIONS}

We analyzed the \g-ray measurements by the \emph{Fermi} LAT in the region of
HB~21, a mixed-morphology SNR. We detect significant emission
($\sim$29$\sigma$ associated with the remnant. The
emission is best modeled by a disk centered at
(l,b)=(88\ddeg62$\pm$0\ddeg05,+4\ddeg79$\pm$0\ddeg06) with a radius
r=1\ddeg19$\pm$0\ddeg06 (statistical uncertainties only), so it is well resolved
by the LAT for energies greater than 1~GeV. The \g-ray emission extends over the
whole area of the non-thermal radio shell, larger than the X-ray emitting
thermal core. The emission in \g-rays may extend beyond the radio
shell in a region rich of interstellar matter in the north western part of the
SNR. Furthermore, the brightest \g-ray emitting region coincides with known
shocked molecular clumps. Both results are suggestive that collisions of
shock-accelerated particles with interstellar matter are responsible for the
observed \g-ray emission.

The spectrum is best modeled by a curved function, indicative of a cutoff or
break in the spectrum of the accelerated particles, typical of middle-aged SNRs
in a dense interstellar environment. The total \g-ray luminosity of HB~21 above
100~MeV is estimated to be $\sim4\times10^{34}$~erg~s$^{-1}$, fainter than other
SNRs interacting with MC detected by the LAT. This can be explained by the lower
mass of the molecular clouds supposed to be in interaction with the remnant. The
\g-ray emission may be dominated either by $\pi^0$ decay due to nuclei or by
Bremsstrahlung from energetic electrons. An IC origin is disfavored because it
would require unrealistically low interstellar densities in order to prevent
Bremsstrahlung to dominate. Based on the most likely values for the ISM
densities over the volume of the remnant, in the hadronic-dominated scenario
accelerated nuclei contribute a total energy of
$\sim$2$\times$10$^{49}$~ergs. In the leptonic-dominated scenario we need
accelerated electrons for $\sim$4$\times$10$^{49}$~ergs, and also nuclei for
$\sim$10$^{49}$~ergs. Such energy densities match those required to make SNRs
the dominant source of the Galactic CRs \cite[e.g.][]{drury12}.
Recently, \cite{reichardt12} performa an analysis on HB~21 using 3.5 years of \emph{Fermi} LAT data. The conclusions are in agreement with those presented in this proceeding if we consider that we adopt a distance from HB~21 of 1.7 kpc (according to \cite{Byu06}) instead of 0.8 kpc adopted in earlier works and based on the assocition with the Cygnus OB7 complex. In particular our values of flux and luminosity of the SNR are in agreement. Finally they assume a density of 60 cm$^{-3}$ which is the maximum density estimation. Insted, we consider both the minimum and the maximum possible desities obtaining that the energy accelerated particles is $\sim0.3-4\times10^{49}$ erg, regardless of whether hadronic or leptonic emission is dominant. Finally, in our work also systematic effects due to interstellar emission model and effective area are taken into account.

\bigskip
\begin{acknowledgments}
The $Fermi$ LAT Collaboration acknowledges support from a number of agencies and institutes for both development and the operation of the LAT as well as scientific data analysis. These include NASA and DOE in the United States, CEA/Irfu and IN2P3/CNRS in France, ASI and INFN in Italy, MEXT, KEK, and JAXA in Japan, and the K.~A.~Wallenberg Foundation, the Swedish Research Council and the National Space Board in Sweden. Additional support from INAF in Italy and CNES in France for science analysis during the operations phase is also gratefully acknowledged.

Work supported by Department of Energy contract DE-AC03-76SF00515.
\end{acknowledgments}

\bigskip 
\bibliographystyle{apsrev}
\bibliography{hb21_references1}

\end{document}